\documentclass[twocolumn,secnumarabic,amssymb, nobibnotes, aps, prd]{revtex4}
\usepackage{graphicx}

\begin{document}
\title{Superconducting quantum interference device without Josephson junctions}
\author{A. A. Burlakov, V. L. Gurtovoi, A. I. Il'in, A. V. Nikulov, and V. A. Tulin}
\affiliation{Institute of Microelectronics Technology and High Purity Materials, Russian Academy of Sciences, 142432 Chernogolovka, Moscow District, RUSSIA.} 
\begin{abstract} A new type of a superconducting quantum interference device (SQUID) based on a single superconducting loop without Josephson junctions and with asymmetric link-up of current leads is proposed. This SQUID offers advantages in simplicity of fabrication and higher sensitivity of magnetic flux. Magnetic field dependence of the critical current in  aluminium rings with asymmetric link-up of current leads has been measured in order to confirm the possibility of making this type of SQUID.
 \end{abstract}

\maketitle

\narrowtext

\section{Introduction}
A superconducting quantum interference device (SQUID) has provided the highest sensitivity of magnetic flux and magnetic field measurements for several decades. It is based on a closed superconducting loop with one (rf SQUID) or two (dc SQUID) Josephson junctions \cite{SQUID}. Available SQUIDs implement \cite{SQUID} the relation $\Delta \varphi  + 2\pi \Phi /\Phi _{0}  = 2\pi n$ between the total phase difference $\Delta \varphi $ at the Josephson junctions and the magnetic flux $\Phi $ through the loop. Here, $\Phi _{0} = 2\pi \hbar /q$ is the flux quantum and $q = 2e$ is the charge of the electron pair. Since the supercurrent through a Josephson junction is $I = I_{c} sin \Delta \varphi $ \cite{Barone}, measurable quantities vary from minimum to maximum values on a scale of $\Phi _{0}/2$. In this work, we study a possibility of making a SQUID without Josephson junctions with a more sharp dependence of measurable parameters on the magnetic flux. The idea is based on the effect discovered by V.A. Little and R.D. Parks \cite{LP1962}. The possibility of using a loop \cite{Fink} or even a cylinder \cite{Zakosarenko} without Josephson junctions as a quantum interferometer was studied both many years ago and recently \cite{Finkler}. A fundamentally new suggestion of this work, as compared to the previous ones, is the use of asymmetric link-up of current leads. Asymmetric connection to the interferometer with two Dayem bridges was implemented in recent work \cite{Russo}. A new point as compared to \cite{Russo} is the use of a sharp change in the average value of the critical current under a change in the quantum number near  $\Phi   = (n+0.5)\Phi _{0}$. 

\section{Possibility of a more sharp magnetic flux dependence of the persistent current}
The idea of a new SQUID is based on an one-valued relation between the quantum number $n$, the magnetic flux $\Phi $ through the loop without Josephson junctions, and the velocity of superconducting pairs
$$\oint_{l}dl v =  \frac{2\pi \hbar }{m}(n -  \frac{\Phi }{\Phi _{0}}) \eqno{(1)}$$ 
or the persistent current $I_{p} = sqn_{s}v$. The current $I_{p}$ circulating in the loop with the cross section $s$ and the density $n_{s}$ of superconducting pairs should be equal $I_{p} = (q2\pi \hbar /ml\overline{(sn_{s})^{-1}})(n -\Phi /\Phi _{0}) = I_{p,A}2(n -\Phi /\Phi _{0})$ due to velocity quantization (1), where $\overline{(sn_{s})^{-1}} = l^{-1} \oint_{l}dl(sn_{s})^{-1}$. The value $\overline{(sn_{s})^{-1}} =(sn_{s})^{-1}$ and $I_{p,A} = sn_{s}q\pi \hbar /ml$ in a homogeneous loop, i.e. when the quantity $sn_{s}$ is constant in its segments. Velocity (1), the persistent current, and the kinetic energy
$$E_{n} = \oint _{l} dl sn _{s} \frac{mv _{n}^{2}}{2} = I _{p,A}\Phi _{0} (n - \frac{\Phi }{\Phi _{0}})^{2}  \eqno{(2)}$$
of superconducting pairs in the loop depend on the magnetic flux $\Phi $ and the quantum number $n$. The absolute value $|I_{p}| = I_{p,A}$ of the persistent current at $|n -\Phi /\Phi _{0}| = 0.5$, which determines the energy difference between permitted states (2), increases with a decrease in temperature: $I_{p,A}(T) = I_{p,A}(0)(1 - T/T_{c})$ \cite{ JETP07De}. At the realistic currents $I_{p, A} = 1 \ \mu A; 10 \ \mu A;  100 \ \mu A$  \cite{ JETP07De}, the energy differences $|E_{n+1}- E_{n}| \approx I_{p,A}\Phi _{0} \approx  2 \ 10^{-21}\ J; I_{p,A}\Phi _{0} \approx  2 \ 10^{-20} \ J; I_{p,A}\Phi _{0} \approx  2 \ 10^{-19} \ J$ at $\Phi  \approx n\Phi _{0}$ correspond to the temperatures $T_{dis} = I_{p,A}\Phi _{0}/k_{B} \approx  150 \ K$; $T_{dis} \approx  1500 \ K$; $T_{dis} \approx  15000 \ K$, respectively. At the transition of at least one segment to the normal state with $n_{s} = 0$, the persistent current $I_{p,A} = q\pi \hbar /ml\overline{(sn_{s})^{-1}}$ and energy (2) vanish. At the reverse transition the quantum number takes the value $n$, with the probability $P_{n} \propto exp -E_{n}/k_{B}T$ given by the laws of statistical physics. Numerous measurements of the critical current \cite{JETP07De,Moshchalkov,JETP07Ju} and other quantities \cite{Burlakov07,Koshnick,Lett2003} confirm the overwhelming probability of lowest energy state (2).

The interval $\Delta \Phi _{e}/\Phi _{0} = \Phi /\Phi _{0}  - (n+0.5)$, in which the probability of the state $n$ varies from $P_{n} \approx  1$ äî $P_{n} \approx  0$, depends on the ratio $I_{p,A}\Phi _{0}/k_{B}T = T_{dis}/T$ because $E_{n+1} - E_{n} \approx  I_{p,A}\Phi _{0}0.5\Delta \Phi _{e}/\Phi _{0}$ in the vicinity of half of the flux quantum. The probability changes by a factor of 10 at $\Delta \Phi _{e}/\Phi _{0} \approx  ln10 k_{B}T/I_{p,A}\Phi _{0} = 2.3 T/T_{dis}$ and by a factor of 100 at $\Delta \Phi _{e}/\Phi _{0} \approx  ln100 k_{B}T/I_{p,A}\Phi _{0} = 4.6 T/T_{dis}$. At the temperature of measurements $T \approx  1 \ K$  the probability $P_{n}$ changes by a factor of 10 within the interval $\Delta \Phi _{e} \approx 0.015\Phi _{0}$; $\Delta \Phi _{e} \approx 0.0015\Phi _{0}$; $\Delta \Phi _{e} \approx 0.00015\Phi _{0}$ in the loops with $I_{p,A} = 1 \ \mu A; 10 \ \mu A; 100 \ \mu A$, respectively. These intervals are much smaller than the magnetic flux interval $\Delta \Phi _{e} \approx 0.5\Phi _{0}$ of the variation of the persistent current in rf and dc SQUIDs.

\section{SUPERCONDUCTING LOOP with asymmetric link-up of current leads}
The distribution of the external current
$$I = qs_{long}n_{s,long} v_{long} - qs_{sh}n_{s,sh} v_{sh}  \eqno{(3)}$$
between the long ($l_{long}$ with the cross section $s_{long}$ and the pair density $n_{s,long}$) and short ($l_{sh}$ with the cross section $s_{sh}$ and the pair density $n_{s,sh}$) segments of the loop $l = l_{long} + l_{sh}$ (Fig. 1) is uniquely determined by quantization condition (1):
$$l_{long} v_{long} +  l_{sh} v_{sh} =  \frac{2\pi \hbar }{m}(n -  \frac{\Phi }{\Phi _{0}}) \eqno{(4)}$$
At equal cross sections $s_{long} = s_{sh} = s$ and pair densities $n_{s,long} = n_{s,sh} = n_{s}$, the velocities
$$v_{sh} =  -\frac{l_{long}}{l}\frac{I_{ext}}{qsn_{s}} + \frac{2\pi \hbar }{ml}(n -  \frac{\Phi }{\Phi _{0}}) \eqno{(5a)}$$
and
$$v_{long} =  \frac{l_{sh}}{l}\frac{I_{ext}}{qsn_{s}} + \frac{2\pi \hbar }{ml}(n -  \frac{\Phi }{\Phi _{0}}) \eqno{(5b)}$$
in the short and long segments, respectively, reach the critical value $|v_{sh}| = v_{c}$ at
$$I_{c} =  \frac{l}{l_{long}}qsn_{s}[v_{c} + \frac{2\pi \hbar }{ml}(n -  \frac{\Phi }{\Phi _{0}})] = I_{c0} +  \frac{l}{l_{long}}I_{p} \eqno{(6a)}$$
and $|v_{long}| = v_{c}$ at
$$I_{c} =  \frac{l}{l_{sh}}qsn_{s}[v_{c} - \frac{2\pi \hbar }{ml}(n -  \frac{\Phi }{\Phi _{0}})] = \frac{l_{long}}{l_{sh}}I_{c0} -  \frac{l}{l_{sh}}I_{p} \eqno{(6b)}$$
The positive direction is taken to be from left to right for $I$ and clockwise for all other quantities. The short segment of the loop is situated at the bottom, see Fig. 1.  

\begin{figure}[]
\includegraphics{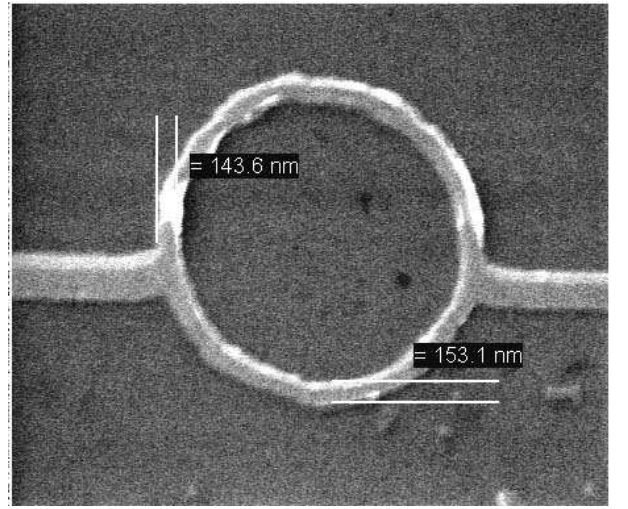}
\caption{\label{fig:epsart}Aluminum ring with the radius $r \approx   1 \ \mu m$ and asymmetric contacts: $l_{long} \approx  1.18\pi r$, $l_{sh} \approx  0.82\pi r$. For this ring, $(l_{long}-l_{sh})l_{long}/l^{2}= 0.11$ .}
\end{figure}

The persistent current $I_{p} = I_{p,A}2(n -\Phi /\Phi _{0})$ jumps from $I_{p} = -I_{p,A}$ to $I_{p} = +I_{p,A}$ with a change in the quantum number $n$ at $\Phi   = (n+0.5)\Phi _{0}$ (see Fig. 4.4 in \cite{Tinkham}). Therefore, critical current (6) must also change abruptly by $(l/l_{long})2I_{p,A}$ or by $(l_{long} - l_{sh})(I_{c0}/l_{sh} - I_{p,A}l/l_{long}l_{sh})$ at a higher $(l_{long} - l_{sh})l_{long}/l^{2} \geq  I_{p,A}/I_{c0}$ or lower $(l_{long} - l_{sh})l_{long}/l^{2} < I_{p,A}/I_{c0}$ asymmetry, respectively. The jump vanishes at the symmetric connection of the contacts, $l_{long} = l_{sh} = l/2$.

\section{MAGNETIC FLUX DEPENDENCE OF THE AVERAGE CRITICAL CURRENT AND VOLTAGE}
In the vicinity of a half integer number of flux quanta, at $|\Delta \Phi _{e}| = |\Phi  - (n+0.5)\Phi _{0}| \ll  \Phi _{0}$, critical current (6a) is $I_{c} \approx  I_{c0} - (l/l_{long})I_{p,A}$ and $I_{c} \approx  I_{c0} + (l/l_{long})I_{p,A}$ in the states $n$ and $n + 1$, respectively. At the dc external current $I \approx  I_{c0}$, the voltage at the ring must be $V = 0$ and $V \approx  R_{n}I_{c0}$ at $n + 1$ and $n$, respectively. The average critical current $\overline{I_{c}} \approx  I_{c0} + (l/l_{long})I_{p,A} (P_{n+1} - P_{n})$ and voltage $\overline{V} \approx  R_{n}I_{c0}P_{n}$ should vary between the minimum and maximum values within the same narrow interval $|\Delta \Phi _{e}|$ of the magnetic flux as the probabilities $P_{n}$ and $P_{n + 1}$ of the states $n$ and $n + 1$. Since the resistance of the loop in the normal state is $R_{n}  = \rho l_{long}l_{sh}/sl$ and $I_{c0} = sj_{c}l/l_{long}$, one has $R_{n}I_{c0} = \rho j_{c}l_{sh}$. The change in the voltage is on the order of $\rho j_{c}l_{sh} \approx  10 \ mV$ at $l_{sh} \approx  1 \ \mu m$ and typical values of the resistivity $\rho \approx  10^{-5} \ \Omega cm$ and critical current density $j_{c} \approx 10^{7} \ A/cm^{2}$ of known superconductors, e.g., niobium. This change is several orders of magnitude greater than the voltage change per flux quantum in a dc SQUID \cite{SQUID}. The sensitivity $\partial \overline{V}/\partial \Phi$ (an important parameter for the use of a dc SQUID as a measuring device \cite{SQUID}) can be additionally increased owing to the variation of the average voltage $\overline{V} \approx  R_{n}I_{c0}P_{n}(\Delta \Phi _{e})$ in a much smaller magnetic flux interval than the flux quantum $\Delta \Phi _{e} \ll  \Phi _{0}$.

To measure the average voltage $\overline{V}$, the loop must be switched to the normal state for a short time. This can be accomplished by short $\Delta t$ current pulses $-\Delta I_{ext} > 2I_{ext}$, the sign of which is opposite to the dc external current $I_{c1} < I_{ext} < I_{c2}$, see Fig. 2. The signs of $I_{ext}$ and $\Delta I_{ext}$ must be opposite owing to the hysteresis of the current-voltage characteristics (Fig. 2). The average voltage measured for a time much longer than the pulse period $T$ should be $\overline{V} \approx R_{n}I_{ext}P_{1}$ at a short pulse length $\Delta t/T \ll 1$. The measured voltage $\overline{V}$ should vary within the same narrow interval of $\Phi $ values as the probability $P_{1}$ of the state with a lower critical current.

\begin{figure}[]
\includegraphics{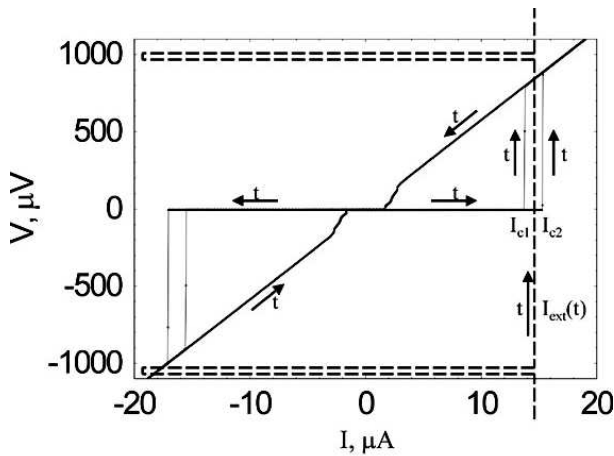}
\caption{\label{fig:epsart} Four typical voltage-current characteristics of an aluminum ring measured at the temperature $T \approx  0.86T_{c} \approx  1.32 \ K$. Two pairs of curves were measured under the variation of the external current $I$ from -20 to 20 $\mu A$ and from 20 to -20 $ \mu A$. Variations of the current and voltage are shown by arrows. Upon the achievement of the critical current ($I_{c1}$ or $I_{c2}$ corresponding to two measurements), the voltage jumps from $V = 0$ to $V = R_{n}I_{c}$. The dashed line shows the variation of the external current with time (the time increases upward) required for the measurement of the time average voltage  $\overline{V} \approx R_{n}I_{ext}P_{1}$, which is proportional to the probability $P_{1}$ of the state with the lower critical current $I_{c} = I_{c1}$.}
\end{figure}

\section{VOLTAGE-CURRENT CHARACTERISTICS AND MAGNETIC FLUX DEPENDENCE OF THE CRITICAL CURRENT IN ALUMINUM RINGs with asymmetric link-up of current leads}
To confirm the theoretical predictions, we performed the first measurements of the magnetic field dependence of the critical current in the aluminum ring with the asymmetric connection of contacts. The structure shown in Fig. 1 was fabricated from an aluminum film with the thickness $d = 20 \ nm$ on a $Si/SiO_{2}$ substrate by electron lithography with the use of liftoff technique. The electron exposure of the pattern (electron lithography) was performed on an EVO 50 scanning electron microscope equipped with a Nano Maker software/hardware system. The radius and width of the ring were $r \approx   1 \ \mu m$ and $w \approx   0.15 \ \mu m$, respectively. The resistance of the ring in the normal state was $R_{n} \approx  60 \ \Omega $. The resistance ratio was $R(300K)/R(4.2K) \approx  1.6$. The superconducting transition temperature was $T_{c} \approx  1.52 \ K$. The measurements were carried out by a fourterminal method in a glass helium cryostat. We used 4He as the cooling agent and pumping of helium allowed lowering the temperature down to 1.19 K. The temperature was measured by a calibrated thermistor ($R(300K)=1.5 \ k\Omega $) with an excitation current of $0.1 \ \mu A$.

The dependences $I_{ñ+}(B)$ and $I_{ñ-}(B)$ of the critical current on the magnetic field were found from the periodic (10 Hz) current-voltage characteristics (see Fig. 2) in a slowly ($\sim 0.01$ Hz) varying magnetic field $B_{sol}$ according to the following algorithm: (i) the superconducting state of the structure was verified; (ii) after the threshold voltage (set above the pickup and noise level of the measuring circuit and determining the lowest measurable critical current) was exceeded, the magnetic field and critical current were measured with a delay of about $30 \ \mu s$. Thus, the critical current in the positive ($I_{c+}$) and negative ($I_{c-}$) directions of the external current $I$ were measured in sequence. Recording one $I_{ñ+}(B)$, $I_{ñ-}(B)$ curve of 1000 data points took 100 s. The magnetic field $B$ perpendicular to the sample plane was produced by a copper coil situated outside the cryostat. The measured quantities were recorded as functions of the current $I_{sol}$ in the coil. The magnitude of the magnetic field induced by the current in the coil was determined from the calibration $B_{sol} = k_{sol}I_{sol}$ with $k_{sol} \approx  129 \ G/A$ found with the use of a Hall probe. To reduce the effect of the Earth's magnetic field, the part of the cryostat where the sample was situated was screened by a permalloy cylinder. The measurement of the critical currents in opposite directions allowed us to determine the external magnetic field $B = B_{sol}+B_{res}$. Since simultaneous change of the direction of the total external magnetic field $B$ and the external current is equivalent to the rotation by $180^{0}$, one has $I_{ñ+}(B) = I_{ñ+}(B_{sol}+B_{res}) = I_{ñ-}(-B) = I_{ñ-}(-B_{sol}-B_{res})$. The residual magnetic field $B_{res} \approx  0.1 \ G$ thus determined corresponds to the flux $SB_{res} = \pi r^{2}B_{res} \approx  0.02\Phi _{0}$ in the ring with the radius $r \approx   1 \ \mu m$.

\begin{figure}[b]
\includegraphics{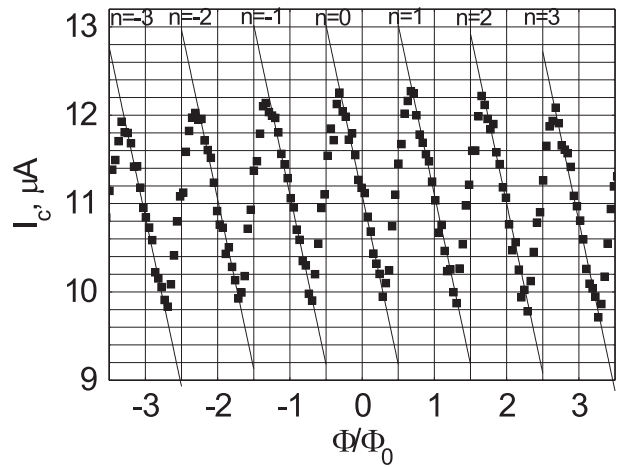}
\caption{\label{fig:epsart} Critical current in the aluminum ring shown in Fig. 1 versus the magnetic flux through the ring measured at the temperature $T \approx  0.94T_{c} \approx  1.44 \ K$. The lines are theoretical curves (6a) corresponding to the critical current $I_{c0}  \approx  11 \ \mu A$ at  $\Phi  = n\Phi _{0}$ and the persistent current amplitude $I_{p,A}  \approx  1.1 \ \mu A$. At  $\Phi  = n\Phi _{0} \neq 0$, $I_{c0}$ decreases (by $\approx 0.2 \mu A$ at  $\Phi  = \pm 3\Phi _{0}$) owing to the suppression of superconductivity by the magnetic field because of the finite width ($w \approx  0.15 \ \mu m$) segments of the structure (Fig. 1).}
\end{figure}

When the current reaches the critical value $I_{c+}$ or $I_{c-}$, the ring switches abruptly to the normal state. In this case, the current-voltage characteristics exhibit hysteresis with a decrease in the external current (see Fig. 2). These features of the current-voltage characteristics typical of one-dimensional superconductors were explained in \cite{ JETP07De}. In this case, it is important that the ring can be switched to the normal state by a short pulse $-\Delta I_{ext} > 2I_{ext}$. After this pulse, the ring will switch back to the superconducting state with a lower ($I_{c1}$) or higher ($I_{c2}$) critical current. In the former and latter cases, the voltage at the external current $I_{c1} < I_{ext} < I_{c2}$ is $V = R_{n}I_{ext}$ and $V = 0$, respectively. The voltage jump in Fig. 2 is  $\Delta V = R_{n}I_{ext} \approx  59 \ \Omega  \times  14 \ \mu A \approx 0.8 \ mV$. The measured value is greater than the product $\rho j_{c}l_{sh} \approx  0.4 \ mV$ of the resistivity $\rho \approx  2 \ 10^{-6} \ \Omega cm$, the critical current density $j_{c} \approx 10^{6} \ A/cm^{2}$ at the temperature $T \approx  0.86T_{c}$, and the length $l_{sh} \approx  2 \ \mu m$ of the short segment owing to the additional resistance of the wires with a length of $\approx 4 \ \mu m$.

The measurements performed at various temperatures $T > 0.85T_{c}$ indicated that the dependences $I_{c+}(\Phi /\Phi _{0})$ (Fig.3) and $I_{c-}(\Phi /\Phi _{0})$ agree with theory (6) near the integer number of flux quanta. Oscillations with the period $B_{0} = \Phi _{0}/S \approx  5.66 \ G$ corresponding to the area of the ring with the radius $r \approx 1.08 \ \mu m$ were observed from $B = -15B_{0}$ to $B = +15B_{0}$. Owing to a finite width ($w \approx  0.15 \ \mu m$) of the superconductor forming the ring, the critical current decreased smoothly with an increase in the magnetic field $B$ (Fig. 3). The values $I_{c0} \approx  11 \ \mu A$ and $I_{p,A} \approx  1.1 \ \mu A$ correspond to a higher asymmetry $(l_{long} - l_{sh})l_{long}/l^{2} \approx  0.11 \geq  I_{p,A}/I_{c0} \approx  0.10$ of the ring with $l_{long} \approx 1.18\pi r$ and $l_{sh} \approx 0.82\pi r$ (see Fig. 1). According to the above estimates, the probability $P_{n}$ of the state $n$ at the measured current $I_{p,A} \approx  1.1 \ \mu A$ should change by a factor of 10 within the interval $\Delta \Phi \approx  0.014 \Phi _{0}$ and the derivative $\partial \overline{V}/\partial \Phi  \approx  57 \ mV/\Phi _{0}$ should be much higher than in a dc SQUID. Only the dependences $\overline{V}(\Phi /\Phi _{0})$ and $\overline{I_{c}}(\Phi /\Phi _{0} )$ of the average values must be continuous. Individual measurements should show the discontinuity of $I_{c+}(\Phi /\Phi _{0})$ and $I_{c-}(\Phi /\Phi _{0})$ at $\Phi  = (n+0.5)\Phi _{0}$ (6a). The values between $I_{c} \approx  I_{c0} - (l/l_{long})I_{p,A}$ (at $n$) and $I_{c} \approx  I_{c0} + (l/l_{long})I_{p,A}$ (at $n + 1$) cannot be observed. However, our measurements showed deviations from theory (6) in the intervals $(n-0.33)\Phi _{0} > \Phi  > (n+0.33)\Phi _{0}$ and the critical current (Fig. 3) contradicts quantization condition (1).

\section{DISCUSSION}
In conclusion, the measurements of the current-voltage characteristics have confirmed the possibility of obtaining a higher derivative $\partial \overline{V}/\partial \Phi  $ with the use of a superconducting loop with asymmetric contacts. The measurement of the magnetic field dependence of the critical current has revealed disagreement with the theoretical predictions, which requires further investigation. We hope that a jump in the critical current associated with a change in the quantum number $n$ will be discovered in the investigations of other structures with asymmetric contacts probably made of other superconductors.


\begin{thebibliography}{99}

\bibitem{SQUID} Superconductor Applications: SQUIDs and Machines, Ed. by B. B. Schwartz and S. Foner (Plenum, New York, 1977).

\bibitem{Barone} A. Barone and G. Paterno, Physics and Applications of the Josephson Effect (Wiley, New York, 1982).

\bibitem{LP1962} W. A. Little and R. D. Parks, {\em Phys. Rev. Lett}. {\bf 9}, 9 (1962).

\bibitem{Fink} H. J. Fink, V. Grunfeld, and A. Lopez, {\em Phys. Rev. B} {\bf 35}, 35 (1987).

\bibitem{Zakosarenko} V. M. Zakosarenko, E. V. Il'ichev, and V. A. Tulin, {\em Sov. Tech. Phys. Lett}. {\bf 15}, 17 (1989).

\bibitem{Finkler} A. Finkler, Y. Segev, Y. Myasoedov, M. L. Rappaport, L. Neeman, D. Vasyukov, E. Zeldov, M. E. Huber, J. Martin, and A. Yacoby, {\em Nano Lett}. {\bf 10}, 1046 (2010).

\bibitem{Russo} R. Russo, C. Granata, E. Esposito, D. Peddis, C. Cannas, and A. Vettoliere, {\em Appl. Phys. Lett}. {\bf 101}, 122601 (2012).

\bibitem{JETP07De} V. L. Gurtovoi, S. V. Dubonos, A. V. Nikulov, et al., {\em Zh. Eksp. Teor. Fiz}. {\bf 132}, 1320 (2007) [{\em JETP} {\bf 105}, 1157 (2007)].

\bibitem{Moshchalkov} D. S. Golubovis and V. V. Moshchalkov, {\em Appl. Phys. Lett.} {\bf 87}, 142501 (2005). 

\bibitem{JETP07Ju} V. L. Gurtovoi, S. V. Dubonos, S. V. Karpii, et al., {\em Zh. Eksp. Teor. Fiz}. {\bf 132}, 297 (2007) [{\em JETP} {\bf 105}, 262 (2007)].

\bibitem{Burlakov07} A. A. Burlakov, V. L. Gurtovoi, S. V. Dubonos, A. V. Nikulov, and V. A. Tulin, {\em JETP Lett.} {\bf 86}, 517 (2007).

\bibitem{Koshnick} N. C. Koshnick, H. Bluhm, M. E. Huber, and K. A. Moler, {\em Science} {\bf 318}, 1440 (2007). 

\bibitem{Lett2003} S. V. Dubonos, V. I. Kuznetsov, I. Zhilyaev, et al., {\em Pis'ma Zh. Eksp. Teor. Fiz.} {\bf 77}, 439 (2003) [{\em JETP Lett.} {\bf 77}, 371 (2003)].

\bibitem{Tinkham}  M. Tinkham, {\em Introduction to Superconductivity} (McGraw- Hill, New York, 1975; Atomizdat, Moscow, 1980).

\end{thebibliography}
\end{document}